\documentclass[letterpaper,english,aps,preprint]{revtex4}
\usepackage[T1]{fontenc}
\usepackage[latin1]{inputenc}
\usepackage{amssymb}

\makeatletter

\makeatletter



\usepackage{babel}
\makeatother

\usepackage{babel}
\makeatother
\begin{document}

\title{Generalized Relativistic Chapman-Enskog Solution of the Boltzmann
Equation}

\author{A. L. García-Perciante$^{1}$, A. Sandoval-Villalbazo$^{2}$, L.
S. García-Colín$^{3}$}

\address{$^{1}$Depto. de Matemáticas Aplicadas y Sistemas, Universidad Autónoma
Metropolitana-Cuajimalpa, Artificios \#40, México DF 01120, México.\\
 $^{2}$Depto. de Física y Matemáticas, Universidad Iberoamericana,
Prolongación Paseo de la Reforma 880, México D. F. 01210, México\\
 $^{3}$Depto. de Física, Universidad Autónoma Metropolitana-Iztapalapa,
Av. Purísima y Michoacán S/N, México D. F. 09340, México. Also at
El Colegio Nacional, Luis González Obregón 23, Centro Histórico, México
D. F. 06020, México. }

\date{\today{}}

\begin{abstract}
The Chapman-Enskog method of solution of the relativistic Boltzmann
equation is generalized in order to admit a time-derivative term associated
to a thermodynamic force in its first order solution. Both existence
and uniqueness of such a solution are proved based on the standard
theory of integral equations. The mathematical implications of the
generalization here introduced are thoroughly discussed regarding
the nature of heat as chaotic energy transfer in the context of relativity
theory.
\end{abstract}
\maketitle

\section{Introduction}

The standard treatment of the Boltzmann equation (BE) provides microscopic
support to the transport equations commonly used in science and engineering
problems \cite{CC}\cite{FK}. It is well known that whereas the local
equilibrium solution to the homogeneous BE leads to the Euler equations
of hydrodynamics, its first order correction in the Knudsen parameter
yields the Navier-Stokes-Fourier transport equations of fluid mechanics
for a dilute gas. In the standard non-relativistic method of establishing
the first order correction to the local equilibrium function, all
partial time derivatives are replaced with spatial gradients using
the Euler equations. Both existence and uniqueness of the solutions
built in this scheme have been shown using the theory of integral
equations \cite{CH}.

On the other hand, in relativity theory time is simply a new coordinate
$ct=x^{4}$. According to this fact, the direct application of the
conventional method used to obtain the first order correction to the
local equilibrium function is shown here to lead to a first order
correction to the distribution function in terms of both spatial and
time derivatives of the local variables namely, the generalized four-component
thermodynamic forces. This is done following the tenets of relativistic
linear irreversible thermodynamics as shown in Ref. \cite{NOS2}.
We wish to emphasize that in this work the Meixner version of non-equilibrium
thermodynamics is developed. However, in most standard works on relativistic
kinetic theory \cite{CK,DEGROOTREL,LIBOFF,ECKART} these ideas are
ignored. The consequence of the absence of time components in the
structure of the thermodynamic forces is on the one hand, the obtention
of transport equations which apparently violate causality, and on
the other hand the tensorial structure of the heat flux in relativity,
which is still a subject of debate \cite{SRM}. \emph{The main motivation
of this work is precisely to critically analyze these fine points.}

The mathematical procedure here used to show that the solution obtained,
considering four component thermodynamic forces, exists and is unique
is based on the standard theory of integral equations. One requires
orthogonality between the inhomogeneous part of the equation and the
collisional invariants. This relation implies the validity of Euler
equations. Nevertheless the use of these equations is not required
for the construction of the solution to first order in the gradients.
These results have strong implications on the structure of the four
component heat flux as will be discussed in this work. Finally, the
usual subsidiary conditions are invoked in order to adjust integration
constants. This has, as will be discussed in this work, strong implications
on the structure of the heat flux tensor.

The paper is divided as follows. In Section 2 the relativistic BE
and some elements of the Chapman-Enskog method of solution are briefly
reviewed. The general form of the first order in the gradients solution
to the relativistic BE is proposed in Section 3 and both its existence
and uniqueness are proved in Section 4. Finally, Section 5 includes
a discussion of the implications of the results and some concluding
remarks.

\section{The Chapman-Enskog expansion }

The starting point of the generalized formalism is the special relativistic
Boltzmann equation for a simple system in the absence of external
forces \cite{DEGROOTREL,CK}:

\begin{equation}
v^{\alpha}f_{,\alpha}=J\left(f,\, f'\right)\,,\label{eq:1}\end{equation}

\noindent which describes the evolution of the single particle distribution
function $f=f(x^{\nu},\, v^{\nu}|\, n(x^{\nu},t),\, u(x^{\nu},t),\,\varepsilon(x^{\nu},t))$
namely, the molecular number density in phase space. In Eq. (\ref{eq:1})

\noindent \begin{equation}
f_{,\alpha}=\left[\begin{array}{c}
\frac{\partial f}{\partial x^{k}}\\
\frac{1}{c}\frac{\partial f}{\partial t}\end{array}\right]\,,\label{eq:1.1}\end{equation}
 and the index $\alpha$, as well as all greek indices in the rest
of this work, runs from 1 to 4 while the latin ones run up to 3. The
local variables, number density $n$, hydrodynamic velocity $u^{k}$
and internal energy $\varepsilon$ per unit of mass are defined in
terms of $f$ as follows

\noindent \begin{equation}
n=\int f\gamma dv^{*}\,,\label{eq:01}\end{equation}

\noindent \begin{equation}
nu^{k}=\int f\gamma v^{k}dv^{*}\,,\label{eq:02}\end{equation}

\noindent \begin{equation}
n\varepsilon=mc\int f\gamma v^{4}dv^{*}\,.\label{eq:03}\end{equation}

\noindent Here $v^{\mu}$ is the molecular velocity four-vector which
we denote in terms of the three velocity $\vec{w}$ as

\noindent \begin{equation}
v^{\alpha}=\left[\begin{array}{c}
\gamma\vec{w}\\
\gamma c\end{array}\right]\,,\label{eq:04}\end{equation}

\noindent where $\gamma=\left(1-w^{2}/c^{2}\right)^{-1/2}$ is the
usual relativistic factor. Also, the differential velocity element
can be written as \cite{LIBOFF}

\noindent \begin{equation}
dv^{*}=\gamma^{5}\frac{cd^{3}w}{v^{4}}=4\pi c^{3}\sqrt{\gamma^{2}-1}d\gamma\,,\label{eq:05}\end{equation}

\noindent and, as usual, $c$ is the speed of light. Equation (\ref{eq:1})
implies that changes in the distribution function are due to collisions,
represented on the right side of Boltzmann's equation through the
so called collision kernel $J\left(f,\, f'\right)$ which is given
by

\noindent \begin{equation}
J\left(f,\, f'\right)=\int\int\left(f'\, f'_{1}-f\, f_{1}\right)\sigma\left(\Omega\right)gd\Omega dv_{1}^{*}\,.\label{eq:2}\end{equation}

\noindent In Eq. (\ref{eq:2}) primes denote quantities after a binary
collision between particles with velocity $v$ and $v_{1}$ takes
place; $\sigma\left(\Omega\right)d\Omega$ is the differential cross
section element and $g$ is the relative velocity.

\noindent To solve Eq. (\ref{eq:1}) following what is now known in
the literature as the Chapman-Enskog method \cite{CC}, originally
due to Hilbert, the distribution function is expanded in a power series
of the Knudsen's parameter $\epsilon$ around the local equilibrium
distribution function $f^{(0)}$

\begin{equation}
f=f^{\left(0\right)}\left(1+\epsilon\phi^{\left(1\right)}+\epsilon^{2}\phi^{\left(2\right)}+\cdots\right)\,.\label{eq:3}\end{equation}

The expansion parameter $\epsilon$ is a measure of the relative magnitude
of the gradients of the local variables within a mean free path and
the characteristic size of the system. The local equilibrium distribution
function, solution to the homogeneous relativistic BE $J\left(f,\, f'\right)=0$,
is the well known Jüttner function. For particles of rest mass $m$,
relativistic parameter $z=\frac{kT}{mc^{2}}$ and in the non-degenerate
case \cite{CK,DEGROOTREL}:

\begin{equation}
f^{\left(0\right)}=\frac{n}{4\pi c^{3}z\mathcal{K}_{2}\left(\frac{1}{z}\right)}e^{-\frac{\gamma}{z}}\,.\label{eq:4}\end{equation}

\noindent As usual, $k$ is the Boltzmann constant, $n$ and $T$
are the local density and temperature, respectively, and $\mathcal{K}_{2}$
represents the modified Bessel function of the second kind.

Substitution of Eq. (\ref{eq:3}) in Eq. (\ref{eq:1}) and considering
only linear deviations in $\epsilon$ from local equilibrium yields

\noindent \begin{equation}
v^{\alpha}f_{,\alpha}^{\left(0\right)}=J\left(f^{\left(0\right)}\left(1+\epsilon\phi\right),\, f'^{\left(0\right)}\left(1+\epsilon\phi'\right)\right)\,,\label{eq:5}\end{equation}

\noindent where $\epsilon\phi^{\left(1\right)}\equiv\phi$ and $f^{\left(0\right)}$,
as pointed above, is the solution to the homogeneous equation namely,\[
J\left(f^{\left(0\right)},\, f^{\left(0\right)}\right)=0\,,\]
or\begin{equation}
f^{\left(0\right)}f'^{\left(0\right)}=f^{\left(0\right)}f_{1}^{\left(0\right)}\,.\label{eq5.5}\end{equation}

\noindent Introducing the relation given by Eq. (\ref{eq5.5}) in
the right hand side of Eq. (\ref{eq:5}) leads to

\noindent \begin{equation}
J\left(f,\, f'\right)=f^{\left(0\right)}\int\int\left[\phi'_{1}+\phi'-\phi_{1}-\phi\right]f_{1}^{\left(0\right)}\sigma\left(\Omega\right)gd\Omega dv_{1}^{*}\,.\label{eq:6}\end{equation}

\noindent Thus, this procedure yields a linear inhomogeneous integral
equation for $\phi$, namely

\noindent \begin{equation}
v^{\alpha}f_{,\alpha}^{(0)}=f^{\left(0\right)}\mathbb{C}\left(\phi\right)\,,\label{eq:57}\end{equation}

\noindent where

\noindent \begin{equation}
\mathbb{C}\left(\phi\right)=\int\int\left\{ \phi'_{1}+\phi'-\phi_{1}-\phi\right\} f_{1}^{\left(0\right)}\sigma\left(\Omega\right)gd\Omega dv_{1}^{*}\,,\label{eq:6}\end{equation}

\noindent is the linearized Boltzmann collision kernel. The following
section is devoted to the obtention of the solution to Eq. (\ref{eq:57}).

\section{Solution to first order in the gradients}

\noindent Following the theory of integral equations as presented
in Ref. \cite{CH}, the general solution to Eq. (\ref{eq:57}) may
be written as the sum of a solution to the inhomogeneous equation
$\phi_{P}$ and an arbitrary linear combination of the solutions of
the homogeneous equation $\phi_{H}$:

\noindent \begin{equation}
\phi=\phi_{H}+\phi_{P}\,.\label{eq:7}\end{equation}

\noindent The homogeneous equation has five solutions, the collisional
invariants namely,

\noindent \begin{equation}
\mathbb{C}\left(\begin{array}{c}
mv^{\mu}\\
m\gamma v^{4}\end{array}\right)=0\,.\label{eq:8}\end{equation}
 Notice that, for $\mu=4$ Eq. (\ref{eq:8}) corresponds to the particle
number conservation.

\noindent A solution given by Eq. (\ref{eq:7}) exists if these invariants
in $\phi_{H}$ are orthogonal to the left hand side of Eq. (\ref{eq:57})
namely

\noindent \begin{equation}
\int\left(\begin{array}{c}
mv^{\mu}\\
m\gamma v^{4}\end{array}\right)v^{\alpha}f_{,\alpha}^{(0)}dv^{*}=0\,.\label{eq:9}\end{equation}

\noindent The relations in Eq. (\ref{eq:9}) correspond precisely
to the relativistic Euler equations and thus the orthogonality conditions
are satisfied. However we recall that there is an infinite number
of solutions since $\phi_{H}$ is an arbitrary linear combination
of collisional invariants. The unicity is achieved when the arbitrary
constants are determined with the use of the subsidiary conditions
as we shall argue below.

In order to find a particular solution to the inhomogeneous equation,
we start by calculating the left hand side of Eq. (\ref{eq:57}).
In the comoving frame and assuming that no external forces are present

\noindent \begin{equation}
v^{\alpha}f_{,\alpha}^{(0)}=v^{\alpha}\left(\frac{\partial f^{\left(0\right)}}{\partial n}n_{,\alpha}+\frac{\partial f^{\left(0\right)}}{\partial T}T_{,\alpha}\right)\,.\label{eq:10}\end{equation}

\noindent Since $\alpha=1,\,2,\,3,\,4$ each gradient on the right
side of Eq. (\ref{eq:10}) has four components. It is precisely at
this point where our formalism deviates from the standard one \cite{CC,CK,DEGROOTREL}.
In conventional relativistic treatments, one separates time and space
derivatives using a projection operator and introduces Euler's equations
in order to eliminate the time derivatives of the local variables.
However, special relativity is a four dimensional formalism in which
time corresponds to a coordinate. Because of this feature we here
propose that the four-gradients are to be considered with the same
status as the generalized thermodynamic forces \cite{DEGROOTFEN}.
\emph{This is why we do not resort to the Euler equations to eliminate
time derivatives.} Also, in most standard relativistic treatments,
the temperature gradient is written in terms of the pressure gradient
in order to introduce, through the momentum balance equation, an acceleration
term so that an Eckart-type constitutive equation is sustained \cite{CK}.
However, in the Meixner-type scheme here considered, such a substitution
needs not to be carried out since the thermodynamic forces are $T_{,\nu}$
and $n_{,\nu}$ exclusively.

In order to proceed, we first notice that since the calculations will
be performed in the comoving frame, $\frac{\partial u^{\alpha}}{\partial t}=0$
and $p_{,k}=0.$ Thus, using the equation of state for an ideal gas
one can write for the space components of the gradients that

\noindent \begin{equation}
\frac{n_{,k}}{n}=-\frac{T_{,k}}{T}\,.\label{eq:11}\end{equation}

\noindent For the time component of the density gradient, in the comoving
frame, one has

\noindent \begin{equation}
n_{,4}=0\,,\label{eq:12}\end{equation}

\noindent since the continuity equation is satisfied to any order
in $\epsilon$. Introducing Eqs. (\ref{eq:11}) and (\ref{eq:12})
in Eq. (\ref{eq:10}) and calculating explicitly the derivatives of
the Jüttner function, the integral equation for $\phi_{P}$ can be
written as

\noindent \begin{equation}
\frac{v^{k}T_{,k}}{zT}f^{\left(0\right)}\left[\gamma-\mathcal{G}\left(\frac{1}{z}\right)\right]+\frac{v^{4}T_{,4}}{T}f^{\left(0\right)}\left[\frac{1}{z}\left(\gamma-\mathcal{G}\left(\frac{1}{z}\right)\right)+1\right]=f^{\left(0\right)}\mathbb{C}\left(\phi_{P}\right)\,,\label{eq:13}\end{equation}

\noindent where $\mathcal{G}\left(\frac{1}{z}\right)=\mathcal{K}_{3}\left(\frac{1}{z}\right)/\mathcal{K}_{2}\left(\frac{1}{z}\right)$.

\noindent Since the thermodynamic forces present on the left hand
side of Eq. (\ref{eq:13}) are independent, we seek a solution as
a linear combination of such forces:

\noindent \begin{equation}
\phi_{P}=\mathcal{A}^{\nu}\frac{T_{,\nu}}{T}\,,\label{eq:14}\end{equation}

\noindent which is a direct generalization of the structure proposed
in the non-relativistic case. The coefficients $\mathcal{A}^{\nu}$,
which are in general functions of $v^{\alpha}$, $n$ and $T$, are
solutions to

\noindent \begin{equation}
\frac{v^{k}}{z}\left[\gamma-\mathcal{G}\left(\frac{1}{z}\right)\right]=\mathbb{C}\left(\mathcal{A}^{k}\right)\,,\label{eq:15}\end{equation}
 for the first three components and,

\noindent \begin{equation}
v^{4}f^{\left(0\right)}\left[\frac{1}{z}\left(\gamma-\mathcal{G}\left(\frac{1}{z}\right)\right)+1\right]=f^{\left(0\right)}\mathbb{C}\left(\mathcal{A}^{4}\right)\,,\label{eq:16}\end{equation}
 for the fourth component. It now remains to show that one may obtain
a unique solution to Eq. (\ref{eq:57}) by using the subsidiary conditions.
This task will be performed in the following section.

\section{Subsidiary conditions and uniqueness of the solution}

According to the previous discussion, the general solution to the
relativistic BE, Eq. (\ref{eq:1}), to first order in the Knudsen
parameter, has the general form

\noindent \begin{equation}
\phi=\mathcal{A}^{\nu}\frac{T_{,\nu}}{T}+\alpha+\tilde{\alpha}_{\nu}v^{\nu}\,,\label{eq:17}\end{equation}

\noindent where the first term corresponds to the particular solution,
Eq. (\ref{eq:14}), while the remaining ones, as mentioned above,
are an arbitrary linear combination of the solutions to the homogeneous
equation, namely the collisional invariants.

\noindent The components of $\mathcal{A}^{\nu}$ are given by the
solution to Eqs. (\ref{eq:15}) and (\ref{eq:16}). However, in order
to uniquely determine the complete solution $\phi$, the constants
$\alpha$ and $\tilde{\alpha}_{\mu}$ must be determined. This is
usually achieved by arbitrarily assuming that the local macroscopic
variables are determined only from the local equilibrium state. That
is,

\noindent \begin{equation}
\int\phi f^{\left(0\right)}\left(\begin{array}{c}
mv^{\mu}\\
m\gamma v^{4}\end{array}\right)dv^{*}=0\,,\label{eq:18}\end{equation}

\noindent which are the so-called subsidiary conditions \cite{CC,FK}.
Since spatial and temporal terms in $\phi_{P}$ have opposite parities
(in $v$), it is convenient at this point to explicitly separate such
terms by proposing that

\noindent \begin{equation}
\mathcal{A}^{\nu}=\left[\begin{array}{c}
a'\left(\gamma\right)v^{k}\\
a''\left(\gamma\right)v^{4}\end{array}\right]\,.\label{eq:19}\end{equation}

\noindent Thus, Eq. (\ref{eq:18}) can be written as

\noindent \begin{equation}
\int\left(a'\left(\gamma\right)v^{k}T_{,k}+a''\left(\gamma\right)v^{4}T_{,4}+\alpha+\tilde{\alpha}_{\nu}v^{\nu}\right)\psi f^{\left(0\right)}dv^{*}=0\,,\label{eq:20}\end{equation}

\noindent where $\psi$ are the collisional invariants $v^{\mu}$
and $\gamma v^{4}$. For $\psi=v^{4}$, the odd parity in $a'\left(\gamma\right)v^{k}T_{,k}$
and $\tilde{\alpha}_{k}v^{k}$ yields zero values for the respective
integrals and thus

\noindent \begin{equation}
\int\left(a''\left(\gamma\right)v^{4}T_{,4}+\alpha+\tilde{\alpha}_{4}v^{4}\right)\gamma f^{\left(0\right)}dv^{*}=0\,.\label{eq:21}\end{equation}

\noindent In a similar way, for $\psi=v^{\ell}$ the only non-vanishing
terms due to parity are $a'\left(\gamma\right)v^{k}T_{,k}$ and $\tilde{\alpha}_{k}v^{k}$,
whence

\noindent \begin{equation}
\int\left(a'\left(\gamma\right)v^{k}T_{,k}+\tilde{\alpha}_{k}v^{k}\right)v^{\ell}f^{\left(0\right)}dv^{*}=0\,,\label{eq:22}\end{equation}

\noindent which is non-zero only for $\ell=k$. In this case

\noindent \begin{equation}
\int\left(a'\left(\gamma\right)v^{k}T_{,k}+\tilde{\alpha}_{k}v^{k}\right)v^{\ell}f^{\left(0\right)}dv^{*}=\frac{1}{3}\int\left(a'\left(\gamma\right)T_{,k}+\tilde{\alpha}_{k}\right)\gamma^{2}w^{2}f^{\left(0\right)}dv^{*}\label{eq:23}\end{equation}

\noindent Thus, the three subsidiary conditions for $\psi=v^{\ell}$
may be written as

\noindent \begin{equation}
\int\left(a'\left(\gamma\right)T_{,k}+\tilde{\alpha}_{k}\right)\gamma^{2}w^{2}f^{\left(0\right)}dv^{*}=0\,.\label{eq:24}\end{equation}

\noindent while the subsidiary condition corresponding to the fifth
collisional invariant, $\psi=\gamma v^{4}$, is

\noindent \begin{equation}
\int\left(a''\left(\gamma\right)v^{4}T_{,4}+\alpha+\tilde{\alpha}_{4}v^{4}\right)\gamma v^{4}f^{\left(0\right)}dv^{*}=0\,.\label{eq:25}\end{equation}

\noindent From Eq. (\ref{eq:24}), it is straightforward to obtain
the relation

\noindent \begin{equation}
\tilde{\alpha}_{k}=-T_{,k}\frac{\int a'\left(\gamma\right)\gamma^{2}w^{2}f^{\left(0\right)}dv^{*}}{\int\gamma^{2}w^{2}f^{\left(0\right)}dv^{*}}\,,\label{eq:26}\end{equation}

\noindent which implies $\tilde{\alpha}_{k}\propto T_{,k}$. Thus,
one can redefine the coefficients appearing in Eq. (\ref{eq:20})
as follows

\noindent \begin{equation}
a'\left(\gamma\right)v^{k}T_{,k}+\tilde{\alpha}_{k}v^{k}\longrightarrow a'\left(\gamma\right)v^{k}T_{,k}\,,\label{eq:27}\end{equation}

\noindent This implies that the subsidiary condition given in Eq.
(\ref{eq:22}) now reads

\noindent \begin{equation}
\int a'\left(\gamma\right)\gamma^{2}w^{2}f^{\left(0\right)}dv^{*}=0\,.\label{eq:28}\end{equation}

\noindent On the other hand, from Eqs. (\ref{eq:21}) and (\ref{eq:25})
an \emph{inhomogeneous} system of equations for $\alpha$ and $\tilde{\alpha}_{4}$
is obtained:

\noindent \begin{equation}
g_{11}\alpha+g_{12}\tilde{\alpha}_{4}=s_{1}T_{,4}\,,\label{eq:29}\end{equation}

\noindent \begin{equation}
g_{21}\alpha+g_{22}\tilde{\alpha}_{4}=s_{2}T_{,4}\,,\label{eq:30}\end{equation}

\noindent where the coefficients are

\noindent \begin{equation}
g_{11}=\int\gamma f^{\left(0\right)}dv^{*}\qquad g_{12}=g_{21}=\int\gamma^{2}cf^{\left(0\right)}dv^{*}\qquad g_{22}=\int c^{2}\gamma^{3}f^{\left(0\right)}dv^{*}\,,\label{eq:31}\end{equation}

\noindent and the coefficients in the inhomogeneous terms are given
by

\noindent \begin{equation}
s_{1}=-\int a''\left(\gamma\right)\gamma^{2}cf^{\left(0\right)}dv^{*}\qquad s_{2}=-\int a''\left(\gamma\right)c^{2}\gamma^{3}f^{\left(0\right)}dv^{*}\,,\label{eq:32}\end{equation}
 The solution to this system of equations is\begin{equation}
\alpha=T_{,4}\frac{s_{1}g_{22}-s_{2}g_{12}}{g_{11}g_{22}-g_{12}g_{21}}\,,\label{eq:33}\end{equation}

\noindent \begin{equation}
\tilde{\alpha}_{4}=T_{,4}\frac{s_{2}g_{11}-s_{1}g_{21}}{g_{11}g_{22}-g_{12}g_{21}}\,.\label{eq:34}\end{equation}
 Since, from Eqs. (\ref{eq:33}) and (\ref{eq:34}), both $\alpha$
and $\tilde{\alpha}_{4}$ are proportional to $T_{,4}$ one can redefine

\noindent \begin{equation}
a''\left(\gamma\right)v^{4}T_{,4}+\alpha+\tilde{\alpha}_{4}v^{4}\longrightarrow a''\left(\gamma\right)v^{4}T_{,4}\,,\label{eq:35}\end{equation}

\noindent and therefore the two corresponding subsidiary conditions
are now given by

\noindent \begin{equation}
\int a''\left(\gamma\right)\gamma^{2}f^{\left(0\right)}dv^{*}=0\,,\label{eq:36}\end{equation}

\noindent \begin{equation}
\int a''\left(\gamma\right)\gamma^{3}f^{\left(0\right)}dv^{*}=0\,,\label{eq:37}\end{equation}

Putting these results together, the unique solution to the relativistic
BE to first order in the gradients according to the Chapman-Enskog
expansion has the general structure

\noindent \begin{equation}
f=f^{\left(0\right)}\left(1+\mathcal{A}^{\nu}\frac{T_{,\nu}}{T}\right)\,,\label{eq:38}\end{equation}

\noindent where $\mathcal{A}^{\nu}$ is defined in Eq. (\ref{eq:19})
and the coefficients $a'\left(\gamma\right)$ and $a''\left(\gamma\right)$
are given by the solution of the integral equations

\noindent \begin{equation}
\frac{v^{k}}{z}f^{\left(0\right)}\left(\gamma-\mathcal{G}\left(\frac{1}{z}\right)\right)=\mathbb{C}\left(a'\left(\gamma\right)v^{k}\right)\,,\label{eq:39}\end{equation}

\noindent \begin{equation}
v^{4}f^{\left(0\right)}\left[\frac{1}{z}\left(\gamma-\mathcal{G}\left(\frac{1}{z}\right)\right)+1\right]=\mathbb{C}\left(a''\left(\gamma\right)v^{4}\right)\,,\label{eq:40}\end{equation}

\noindent subject to the subsidiary conditions given in Eqs. (\ref{eq:28}),
(\ref{eq:36}), and (\ref{eq:37}).

\noindent Notice that, in the non-relativistic case as well as in
the relativistic treatments given in the literature \cite{CK,DEGROOTREL},
the absence of a fourth component of $\mathcal{A}^{\nu}$ leads to
a homogeneous system instead of Eqs. (\ref{eq:29}) and (\ref{eq:30})
and is thus responsible for the appearance of the trivial solutions
$\alpha=\tilde{\alpha}_{4}=0$. The consequences of this substantial
difference are discussed in the next section.

\section{Discussion and Concluding Remarks}

By considering the four-dimensional gradients of local variables as
thermodynamic forces within the Chapman-Enskog method of solution
to the BE, a generalized first correction to the equilibrium distribution
function has been found in the context of Meixner's relativistic irreversible
thermodynamics. In the comoving frame, this procedure leads to a first
order in $\epsilon$ solution with the general structure given by
Eqs. (\ref{eq:38}) and (\ref{eq:19}). The proportionality coefficients,
functions of the local variables and $\gamma$, can be obtained by
an appropriate decomposition in orthogonal functions which simplifies
the integral equations (\ref{eq:39}) and (\ref{eq:40}). Thus, the
solution here obtained is suitable for the calculation of relativistic
transport coefficients consistent with the phenomenological formalism
proposed in Ref. \cite{NOS2}. Such a procedure is beyond the scope
of this work and will be addressed elsewhere.

However one important point should be discussed here, namely the definition
of heat flux in the context of relativity. We recall that in the non-relativistic
theory this quantity is related to the transport of chaotic kinetic
energy in agreement with the interpretation proposed by Clausius in
1857 \cite{brush}. The question here is: can this definition be extended
within a relativistic framework? This, in principle, may be achieved
by resorting to the relativistic version of the Maxwell-Enskog transport
equation. Its explicit form was derived seven years ago \cite{physica2000}
by two of us. This equation, in the notation here introduced reads\begin{equation}
\left(n\left\langle \psi\frac{v^{\alpha}}{\gamma}\right\rangle \right)_{;\alpha}=0\label{eq:ff}\end{equation}
 where $\psi=mv^{\mu},\, m\gamma v^{4}$ is a collisional invariant
and the bracket represents the statistical average defined, for any
function $g\left(\gamma\right)$, as\begin{equation}
\left\langle g\right\rangle =\frac{1}{n}\int fg\gamma dv^{*}\label{eq:gg}\end{equation}
 For $g=mv^{4}$ the heat transport equation is obtained. In particular,
the heat flux will be given by\[
J_{[Q]}^{\nu}=mc^{2}\int v^{\nu}f\gamma dv^{*}\]
 This is a rather striking result because the fourth component of
this vector turns out to be precisely the energy density as defined
in Eq. (\ref{eq:03}). This implies that, because of the subsidiary
conditions, such component will be identically equal to zero for any
perturbation $\phi^{\left(n\right)}$ with $n\geq1$. Thus we recover
the usual form which reduces to the equation obtained by other authors
resorting to the use of a spatial projector \cite{CK,ECKART,DEGROOTREL,ISRAELJMP}.

Indeed to clarify this statement, we recall that Eckart proposed the
following stress-energy tensor \cite{ECKART} for a simple, non-viscous
fluid

\begin{equation}
T_{\mu\nu}=\frac{n\varepsilon}{c^{2}}u_{\mu}u_{\nu}+ph_{\mu\nu}+\frac{1}{c^{2}}u_{\mu}J_{[Q]\nu}+\frac{1}{c^{2}}u_{\nu}J_{[Q]\mu}\label{eq1}\end{equation}
where $h_{\mu\nu}=g_{\mu\nu}-u_{\mu}u_{\nu}/c^{2}$ is the projector
introduced by Eckart in order to eliminate dissipation in the time-axis.
This work thus provides further support for this assumption. However,
$T_{\mu\nu}$ as in Eq. (\ref{eq1}) is in principle coupled to the
Einstein tensor $G_{\mu\nu}$ by means of the field equation: \begin{equation}
G_{\mu\nu}=\mathcal{K}T_{\mu\nu}\label{eq2}\end{equation}
where $\mathcal{K}$ is the coupling constant. The point here is how
heat would affect the gravitational field if it is placed in the stress
energy tensor. Indeed, heat is identified with molecular motion and
its capability of transforming itself to mechanical energy by any
process is restricted. On the other hand, the introduction of $J_{[Q]\nu}$
in Eq. (\ref{eq2}) seems to allow a transformation of heat to work,
via the gravitational field, with unitary efficiency. To the authors'
knowledge, direct attempts to obtain solutions to the system (\ref{eq2})
in the presence of heat and to analyze the dynamical consequences
of the solutions are practically absent in the literature. 

An alternative way of introducing heat in relativistic irreversible
thermodynamics consists in a generalization of Meixner's approach
\cite{NOS2}. In it, heat is not included in the stress energy tensor
but introduced directly in the total energy flux. The construction
of the entropy production, consistent with Clausius' idea of uncompensated
heat, and consequent enforcement of the second law of thermodynamics
in this context leads to the proposal of the constitutive equation\begin{equation}
J_{[Q]}^{\mu}=-\kappa T^{,\mu}\label{eq8}\end{equation}
where $\kappa$ is the thermal conductivity. Equation (\ref{eq8})
coincides with the constitutive equation in Eckart's formalism in
the comoving frame where the acceleration term vanishes.

It is important to emphasize at this point that for the purpose of
this discussion, the key difference between both formalism is the
role of the heat flux. On the one hand it is included in the stress
energy tensor in Eckart's formalism, as pointed out above. On the
other hand, it is conceived only as energy in transit in the Meixner-like
formalism and is introduced separately in the total energy flux. Clearly
the latter conception does not predict any effects on the geometry
of space-time due to heat since $J_{[Q]}^{\mu}$ is not included in
$T_{\nu}^{\mu}$ and thus the Einstein field equation remains unchanged.
An experiment that could clarify this difference has already been
suggested in Ref. \cite{SRM}. 

On the other hand, the validity of Eq. (\ref{eq8}) should be sustained
by solving the relativistic Boltzmann equation. The calculations performed
in the previous sections show that this is not the case. Indeed, the
heat flux has no fourth component which is in agreement with other
theories but for entirely different reasons.

We believe that this poses a rather intriguing question. Is it possible
or not to think of heat as a chaotic molecular motion in a relativistic
framework? The correct answer to this question is crucial when considering
both the structure of transport equations and the consistency of any
relativistic kinetic theory with the second law of thermodynamics
\cite{NOS2}. These questions merit further study.

\end{document}